\documentclass[aps,pra,notitlepage,reprint,superscriptaddress,nofootinbib,longbibliography]{revtex4-2}

\usepackage{graphicx}
\usepackage{dcolumn}
\usepackage{bm}
\usepackage{amsmath}
\usepackage{upgreek}
\usepackage[colorlinks,linkcolor=blue,citecolor=blue,urlcolor=red]{hyperref}
\usepackage[utf8]{inputenc}
\usepackage[T1]{fontenc}
\usepackage{mathptmx}
\usepackage[normalem]{ulem}
\usepackage[colorinlistoftodos]{todonotes}
\usepackage[mathlines]{lineno}
\renewcommand{\t}[1]{\mathrm{#1}}

\DeclareMathAlphabet{\mathcal}{OMS}{cmsy}{m}{n}

\AtBeginDocument{%
    \newwrite\bibnotes
    \def\bibnotesext{Notes.bib}
    \immediate\openout\bibnotes=\jobname\bibnotesext
    \immediate\write\bibnotes{@CONTROL{REVTEX41Control}}
    \immediate\write\bibnotes{@CONTROL{%
    apsrev41Control,author="08",editor="1",pages="1",title="0",year="1"}}
     \if@filesw
     \immediate\write\@auxout{\string\citation{apsrev41Control}}%
    \fi
}%

\begin{document}

\title{Focusing membrane metamirrors for integrated cavity optomechanics}

\author{A. R. Agrawal}%
\affiliation{Wyant College of Optical Sciences, University of Arizona, Tucson, AZ 85721, USA}

\author{J. Manley}
\affiliation{Wyant College of Optical Sciences, University of Arizona, Tucson, AZ 85721, USA}

\author{D. Allepuz-Requena}
\affiliation{Department of Physics, Technical University of Denmark, Kongens Lyngby, Denmark}

\author{D. J. Wilson}
\affiliation{Wyant College of Optical Sciences, University of Arizona, Tucson, AZ 85721, USA}

\date{\today}
\begin{abstract}
We have realized a suspended, high-reflectivity focusing metamirror ($f\approx 10$ cm, $\mathcal{R} \approx 99\%$) by non-periodic photonic crystal patterning of a Si$_3$N$_4$ membrane. The design enables construction of a stable, short ($L$ = 30 $\upmu$m), high-finesse ($\mathcal{F}>600$) membrane cavity optomechanical system using a single plano dielectric end-mirror. 
We present the metamirror design, fabrication process, and characterization of its reflectivity using both free space and cavity-based transmission measurements. The mirror's effective radius of curvature is inferred from the transverse mode spectrum of the cavity. In combination with phononic engineering and metallization, focusing membrane mirrors offer a route towards high-cooperativity, vertically-integrated cavity optomechanical systems with applications ranging from precision force sensing to hybrid quantum transduction.
\end{abstract}

\maketitle


Silicon nitride (Si$_3$N$_4$) membranes have emerged as a leading platform for cavity optomechanics experiments \cite{aspelmeyer2014cavity}, enabling early demonstrations of ground state cooling~\cite{underwood2015measurement,peterson2016laser,rossi2018measurement}, radiation pressure quantum back-action \cite{purdy2013observation}, and ponderomotive squeezing \cite{purdy2013strong,nielsen2017multimode}, and more recent demonstrations such as microwave-optical quantum transduction \cite{andrews2014bidirectional} and entanglement-enhanced force sensing \cite{xia2023entanglement}.  Since the inception of the popular ``membrane-in-the-middle" (MIM) platform \cite{jayich2008dispersive}---in which a membrane is placed between two mirrors forming a Fabry-P\'{e}rot (FP) cavity---a long-standing program has been to engineer membranes with higher $Q$ and higher reflectivity ($\mathcal{R}$), using a combination of phononic (PnC) and photonic (PtC) crystal patterning.  This program has been highly successful, with $Q>10^8$ \cite{tsaturyan2017ultracoherent} and  $\mathcal{R}>99.9\%$ \cite{chen2017high} now routinely achieved (and potentially combined, in recent work \cite{bao2020hybrid, enzian2023phononically}).  Nevertheless, the planar wavefront of the membrane remains an important technical impediment, requiring delicate pre-alignment or active nanopositioning in the MIM topology, or a separate concave mirror in a FP arrangement \cite{bao2020hybrid,enzian2023phononically}.

Here we attempt to overcome the planar wavefront limitation of membrane optomechanics by implementing a high-reflectivity, \textit{focusing} metamirror into a Si$_3$N$_4$ membrane.  Our efforts build on a spate of recent developments in PtC patterning of Si$_3$N$_4$ membranes \cite{bao2020hybrid,enzian2023phononically,zhou2022cavity,norte2016mechanical,chen2017high,bernard2016precision,moura2018centimeter}. 
 Specifically, by combining the hexagonal PtC design of Zhou \textit{et. al.} \cite{zhou2022cavity} with the gradient pitch focusing PtC proposal of Guo \textit{et. al.} \cite{guo2017integrated} we have engineered a $f \approx 10$ cm, $\mathcal{R}\approx 99\%$ (at 850 nm) mirror into a 200 nm thick Si$_3$N$_4$ membrane.  We have also successfully constructed stable, short ($L = 30\,\upmu$m), high-finesse ($\mathcal{F}>600$) cavities using this device by combining it with a plano dielectric mirror.  
In this Letter, we present details on the modeling, fabrication, and characterization of our ``concave'' Si$_3$N$_4$ membrane mirrors and discuss their applicability to optomechanical sensing and quantum experiments.  
In this regard, a long-term motivation for our work is the development of compact, arrayable optomechanical accelerometers~\cite{chowdhury2023membrane} for entanglement-enhanced dark matter searches \cite{manley2021searching,xia2023entanglement,brady2023entanglement}.

\begin{figure}[t!]
\centering
  \vspace{-2pt}
    \includegraphics[width=0.95\columnwidth]{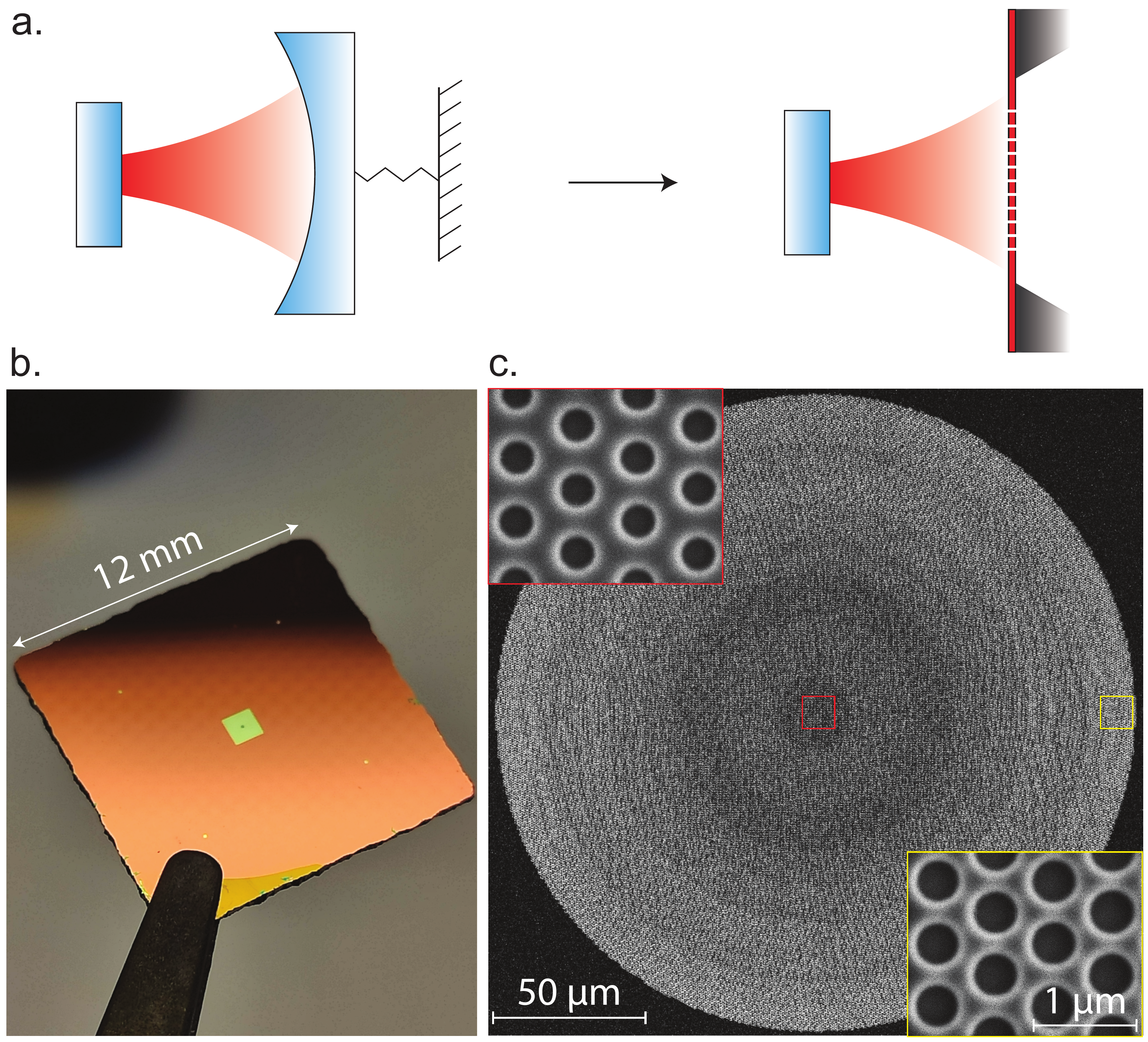}
\caption{(a) Concept for a free-space hemispherical cavity based on a focusing metamirror in a Si$_3$N$_4$ membrane. (b) Camera image of a $1\times 1\,\t{mm}^2$ membrane (on a $12\times 12\,\t{mm}^2$ chip) with a 200 $\upmu$m diameter focusing metamirror etched in its center. (c) SEM image of the device. Insets: magnified image highlighting different period ($a$) and hole radius ($r$) at the mirror center (red box) and edge (yellow box).}
\label{fig:mirrorConcept}
\vspace{-2mm}
\end{figure}  

 \begin{figure*}[t!]
 \vspace{-1mm}
  \centering
  \includegraphics[width=2.0\columnwidth]{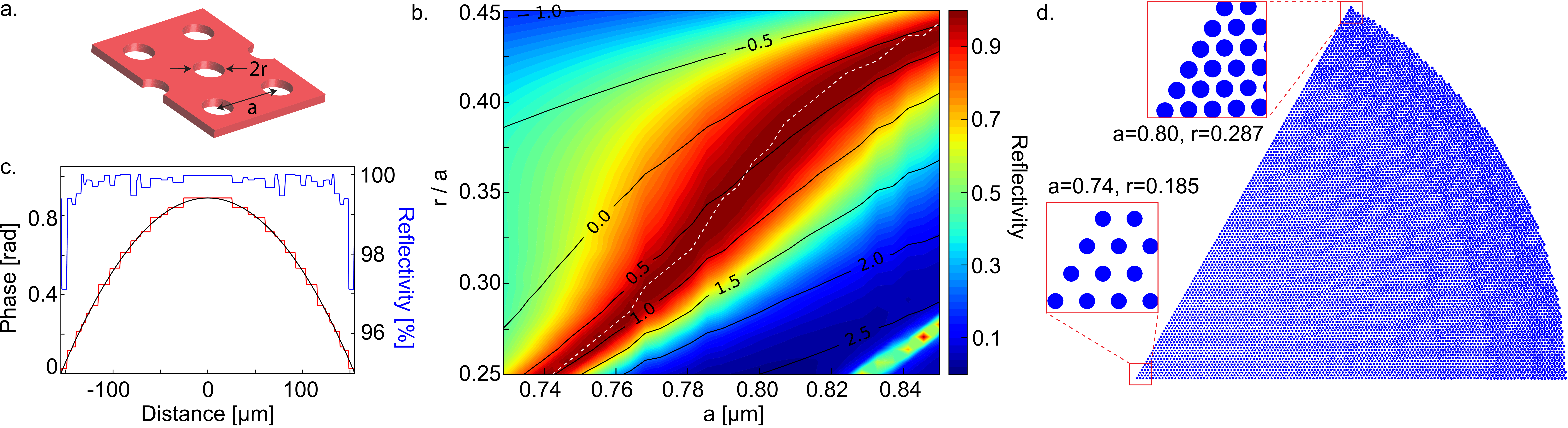}
\caption{Suspended metamirror design. (a) Hexagonal lattice unit cell with period ($a$) and hole radius ($r$). (b) Simulated unit cell reflectivity (color) and phase (black contour lines) versus period and fill factor ($r/a$). The white line indicates the collection of unit cells with reflectivities $>99\%$. (c) Ideal ($\phi_\t{sph}$, black) and design ($\phi_{m,n}$, red) PtC phase profile, constructed from 30 unit cells along the white line in panel b.  The blue line represents the corresponding unit cell reflectivities. (d) Section of the computer-aided PtC design illustrating the unit cell size gradient.}
\label{fig:mirrorDesign}
\vspace{-2mm}
\end{figure*} 

\vspace{-2mm}
 \section{Focusing membrane mirror concept}
\vspace{-2mm}

Figure \ref{fig:mirrorConcept} gives an overview of the focusing membrane mirror concept.  As shown in panel (a), a canonical cavity optomechanical system consists of a Fabry-P\'{e}rot resonator with a mechanically compliant end-mirror.  By replacing the compliant mirror with a membrane, its mass can be reduced; however, in order to maintain the finesse and stability of the cavity, the membrane must be reflective and curved.  To achieve high reflectivity, we pattern a hexagonal PtC into a high-stress Si$_3$N$_4$ membrane (panel b). To achieve curvature, the PtC pattern is varied across the mirror surface (panel c), imparting a spherical phase profile on the reflected wavefront.

 \vspace{-2mm}
 \section{Focusing mirror design}
\vspace{-2mm}

Our focusing metamirror is designed after a traditional spherically concave mirror with a reflective phase profile 
\begin{equation}
\phi_\t{sph}(x,y)=\phi_0-\frac{2\pi}{\lambda}\left(\sqrt{f^2+x^2+y^2}-f\right)   
\end{equation}
where $f$ is the mirror focal length, $(x,y)$ is the coordinate relative to the mirror center $(0,0)$, $\lambda$ is the wavelength of the incident field, and $\phi_0$ is the phase shift at the mirror center. When illuminated at normal incidence by a monochromatic plane wave, a reflector with phase profile $\phi_\t{sph}$ transforms a plane wave into a spherical wave with a radius of curvature $R = 2f$.  When illuminated by a Gaussian beam with a local radius of curvature $R$, the reflected beam is identical, yielding the stability criterion for a two-mirror Fabry-Perot resonator. 

Focusing reflectors are ubiquitous in classical optics and have recently been realized using PtC metasurfaces \cite{fattal2010flat, lu2010planar, ma2015high, cheng2019mechanically, pors2013broadband}. Like all metasurfaces, the basic concept is to use an array of sub-wavelength-spaced optical scatterers (a PtC) to simulate the phase and reflectivity profile of a concave mirror.  Unlike metalenses and planar metamirrors---relatively mature technologies---focusing metamirrors require that the reflectivity and phase of the PtC unit cell be tailored simultaneously and distributed aperiodically over the metasurface. To meet this demand, recent demonstrations focus on high contrast grating \cite{fattal2010flat, lu2010planar, ma2015high} and plasmonic structures \cite{cheng2019mechanically, pors2013broadband}, which feature broadband reflectivity and simple design. Moreover, all these structures were fabricated on thick substrates using additive meta-atoms (grating bars and metal-insulator-metal stacks), enabling $2\pi$ phase coverage and high device yield.


Our aim is to realize a focusing metamirror in a Si$_3$N$_4$ nanomembrane.  Besides delicate handling, the relatively low index ($n \approx 2.0$) of Si$_3$N$_4$ poses a challenge for achieving sufficient phase coverage for small focal lengths.  To mitigate this challenge, we target a relatively large focal length of $f = 10$ cm, compatible with typical membrane-based cavity optomechanical systems employing $10-100\,\upmu\t{m}$ mode diameters.  While both additive (nano-pillar) \cite{norte2016mechanical} and subtractive (nano-hole) \cite{enzian2023phononically,zhou2022cavity,norte2016mechanical,bernard2016precision,moura2018centimeter} PtC membrane mirrors have been demonstrated, we focus on the subtractive design because of its ease of fabrication and correspondingly larger catalog of existing designs.

Figure \ref{fig:mirrorDesign} gives an overview of our PtC design, starting with a circular unit cell of radius $r$ arranged in a hexagonal lattice with pitch $a$ (Fig. \ref{fig:mirrorDesign}a).  In contrast to planar PtC metamirrors, where only the unit cell reflectivity ($\mathcal{R}$) is relevant, designing a focusing metamirror requires mapping both the reflectivity and phase $\phi$ of the unit cell as a function of its geometry (Fig. \ref{fig:mirrorDesign}b). To approximate a spherical phase profile $\phi_\t{sph}(x,y)$, an aperiodic PtC is defined whose unit cell geometry $\{a,r\}$ varies with radial position $\sqrt{x^2+y^2}$, drawing from a contour of approximately constant $\mathcal{R}$ in the reflectivity-phase map (dashed white line in Fig. \ref{fig:mirrorDesign}b).  For a hexagonal lattice, the unit cell positions follow a recurrence relation \cite{guo2017integrated} 
\begin{subequations}
\begin{equation}
    x_{m,n}=x_{m-1,n}+\frac{(a_{m-1,n}+a_{m,n})}{2}
\end{equation}
\vspace{-5.5mm}
\begin{equation}
    y_{m,n}=y_{m,n-1}+{\frac{\sqrt{3}}{2}}\frac{(a_{m,n-1}+a_{m,n})}{2}
\end{equation}
\end{subequations}
where $\{m,n\}=\{0,0\}$ is the index of the central cell, yielding the six-fold-symmetric, graded honeycomb lattice in Fig. \ref{fig:mirrorDesign}d. Here, $a_{m,n}$ is selected to produce a local phase most closely matching the target phase $\left(\phi_{m,n}\approx \phi_\t{sph}(x_{m,n},y_{m,n})\right)$. 

In practice, the metamirror design is constrained by the topology of the unit cell reflectivity-phase map, which depends, in turn, on the thickness and refractive index of the PtC substrate.  In Fig. \ref{fig:mirrorDesign}b, we show unit cell simulations for a 200-nm-thick Si$_3$N$_4$ substrate, using rigorous coupled wave analysis based on the open-source software package, S$^4$ \cite{Liu20122233}.  Monochromatic plane wave incidence ($\lambda = 0.85\,\upmu\t{m}$) and an infinitely periodic base lattice are assumed.  
To scan the design space, we vary the unit cell periodicity $a$ and fill factor $r/a$, ultimately focusing on a region $\{a,r/a\}\in\{[0.74,0.84]\,\upmu\t{m},[0.25,0.45]\}$ over which the reflectivity exhibits a relatively flat ridge-like extremum. The white line in Fig. \ref{fig:mirrorDesign}b---tracing this ridge---represents a set of unit cells with a reflectivity $\mathcal{R} \gtrsim 99$\% and phase range of $\Delta \phi = 0.89\;\t{rad}$, and constitutes a look-up table for the metamirror design.

\begin{figure*}[ht!]
 \vspace{-3mm}
	\centering
	\includegraphics[width=1.75\columnwidth]{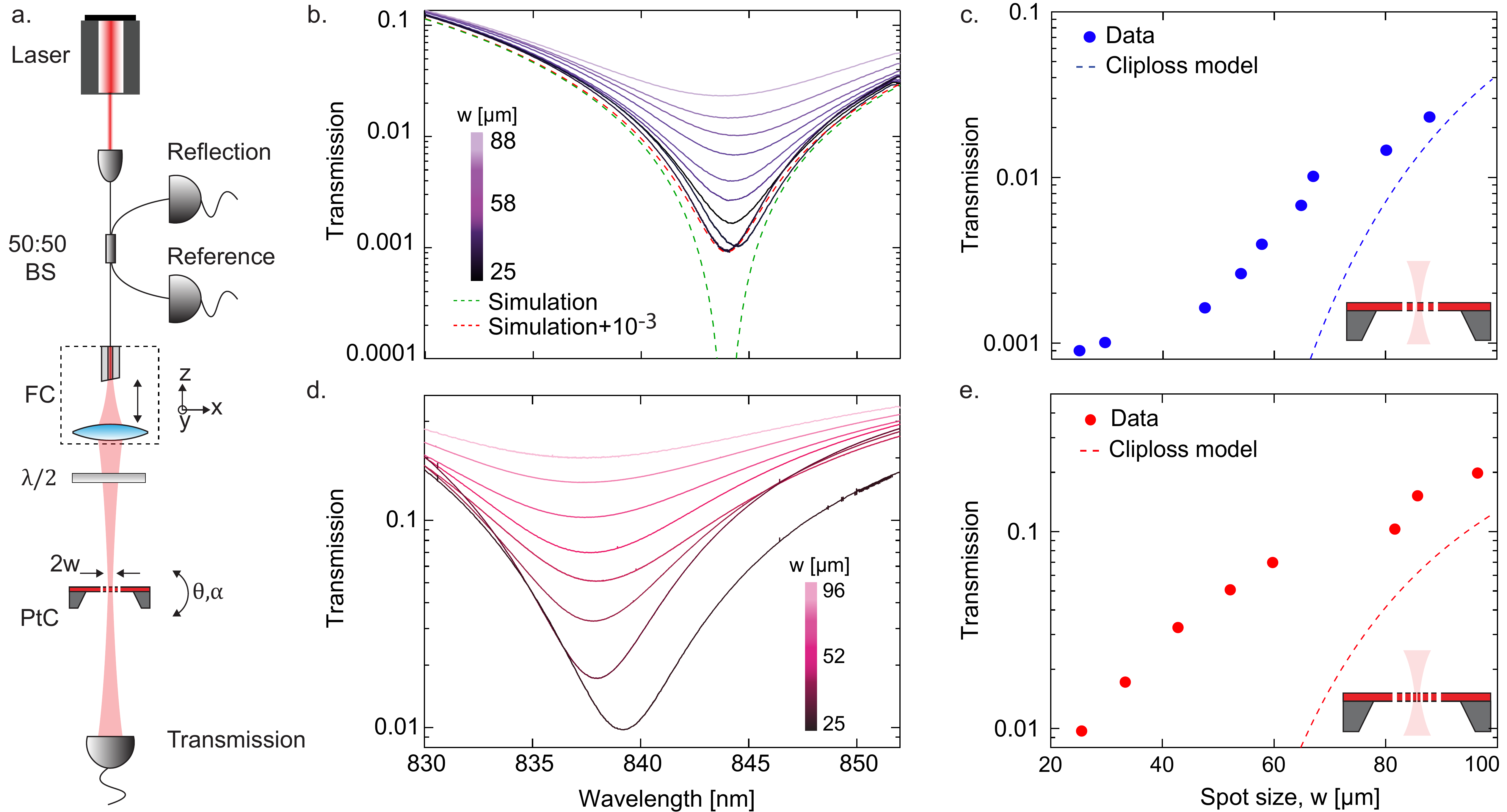}
	\caption{Free-space transmission measurements. (a) Optical setup consisting of fiber-coupled (FC) laser with spot size ($w$) adjustable via the lens-facet separation.  A fiber beamsplitter (BS) provides auxiliary reference and reflection ports for calibrating the input power and aligning the laser to the photonic crystal (PtC) mirror, respectively. (b,d) Measured transmission of plano (b) and focusing (d) PtCs as a function of wavelength and spot size. The dashed green curve is the simulated transmission of a normally incident plane wave; the dashed orange
    curve includes a constant loss of $10^{-3}$. (c,e) Minimum measured transmission of plano (c) and focusing (e) PtCs versus spot size. Dashed lines are clipping loss models for 250-$\upmu$m-diameter plano and 200-$\upmu$m-diameter focusing PtCs, respectively, \mbox{assuming a membrane transmission of $94\%$.}}
    \label{fig:freespaceMeasurements}
 \vspace{-3mm}
\end{figure*}

With the unit cell map in hand, we target a metamirror reflectivity and focal length of $\mathcal{R} = 0.998$ and $f = 10\;\t{cm}$, respectively, corresponding to an aperiodic PtC with $\mathcal{R}_{m,n}\approx \mathcal{R}$ and an approximately parabolic phase profile of $\left(\phi_0-\phi_{m,n}\right) \approx \pi (x_{m,n}^2+y_{m,n}^2)/(\lambda f) = 0.37 (x_{m,n}^2+y_{m,n}^2)/(100\,\upmu\t{m})^2$ (Fig. \ref{fig:mirrorDesign}c).  Our choice of $f$ is informed by several tradeoffs related to the low phase range $\Delta\phi$ of our unit cell. 
In particular, as evident in Fig. \ref{fig:mirrorDesign}d, increasing $f$---and thereby reducing the phase gradient---enables construction of larger mirrors if a constraint is placed on unit cell fill factor at large $\{m,n\}$.  In our case, $f=10\;\t{cm}$ allows fabrication of a $w_\t{m} = 100\,\upmu\t{m}$ radius mirror, limited by the resolution of our electron beam lithographer (see below). This radius is desirable because it permits critically coupled cavities with lengths $L$ and mode radii $w_\t{c}$ as large as 1~mm and 60 $\upmu$m, respectively, subject to the constraint $w_\t{c}\approx (2 L f \lambda^2/\pi^2)^{0.25}\lesssim w_\t{m} \sqrt{2/\ln \left(1-\mathcal{R}\right)^{-1}}$ due to finite aperture clipping loss \cite{kogelnik1966laser}.

\vspace{-2mm}
\section{Metamirror fabrication}
\vspace{-2mm}

The device in Fig. \ref{fig:mirrorConcept}c is a physical realization of the nonperiodic PtC design in Fig. \ref{fig:mirrorDesign}d, etched into a 200-nm-thick Si$_3$N$_4$ membrane. To fabricate this device, we followed a standard hybrid wet-dry etch procedure for subtractive PtC Si$_3$N$_4$ membranes \cite{chen2017high,bernard2016precision}, with some modifications to enable the high unit cell fill factor far from the PtC center. 
The process starts by using photolithography to pattern a square window on the backside of double-sided Si$_3$N$_4$ coated Si chip (WaferPro), followed by wet etching in a potassium hydroxide solution to release a membrane on the front side. The chip is then fixed to a carrier wafer and spin coated with a 350-nm-thick electron beam resist (ZEP 520A). We then pattern the PtC in the center of the membrane using a 100 kV electron beam lithography system (Elionix ELS-7000). After developing the pattern, it is transferred to the suspended membrane using a fluorine-based (CHF$_3$+SF$_6$) slow reactive ion etch. We observe that the slow plasma etch is critical for PtC membrane survival (allowing better heat dissipation and more gradual stress relaxation). Finally, we remove the remaining resist using an NMP (n-Methyl-Pyrrolidone)-based resist stripper (AZ 400T) and rinse the chip in deionized water and isopropanol.

\vspace{-3mm}
\section{Metamirror characterization}
\vspace{-2mm}

Planar Si$_3$N$_4$ membrane mirrors have been characterized using a variety of techniques, including free space scanning wavelength spectroscopy \cite{enzian2023phononically,zhou2022cavity,norte2016mechanical,bernard2016precision,moura2018centimeter}, free space white light spectroscopy \cite{chen2017high}, and cavity-based spectroscopy in the MIM arrangement \cite{chen2017high} or in the FP arrangement with a concave end-mirror \cite{chen2017high,enzian2023phononically}.  To characterize our focusing membrane mirrors, we use two approaches: free-space scanning wavelength spectroscopy, and FP spectroscopy with a \textit{planar} end-mirror.  The ability to form a stable cavity with a planar mirror is a key feature of the focusing metamirror design and manifests as a transverse mode splitting that encodes the mirror focal length.

\begin{figure*}[ht!]
\vspace{-2mm}
  \centering
  \includegraphics[width=1.8\columnwidth]{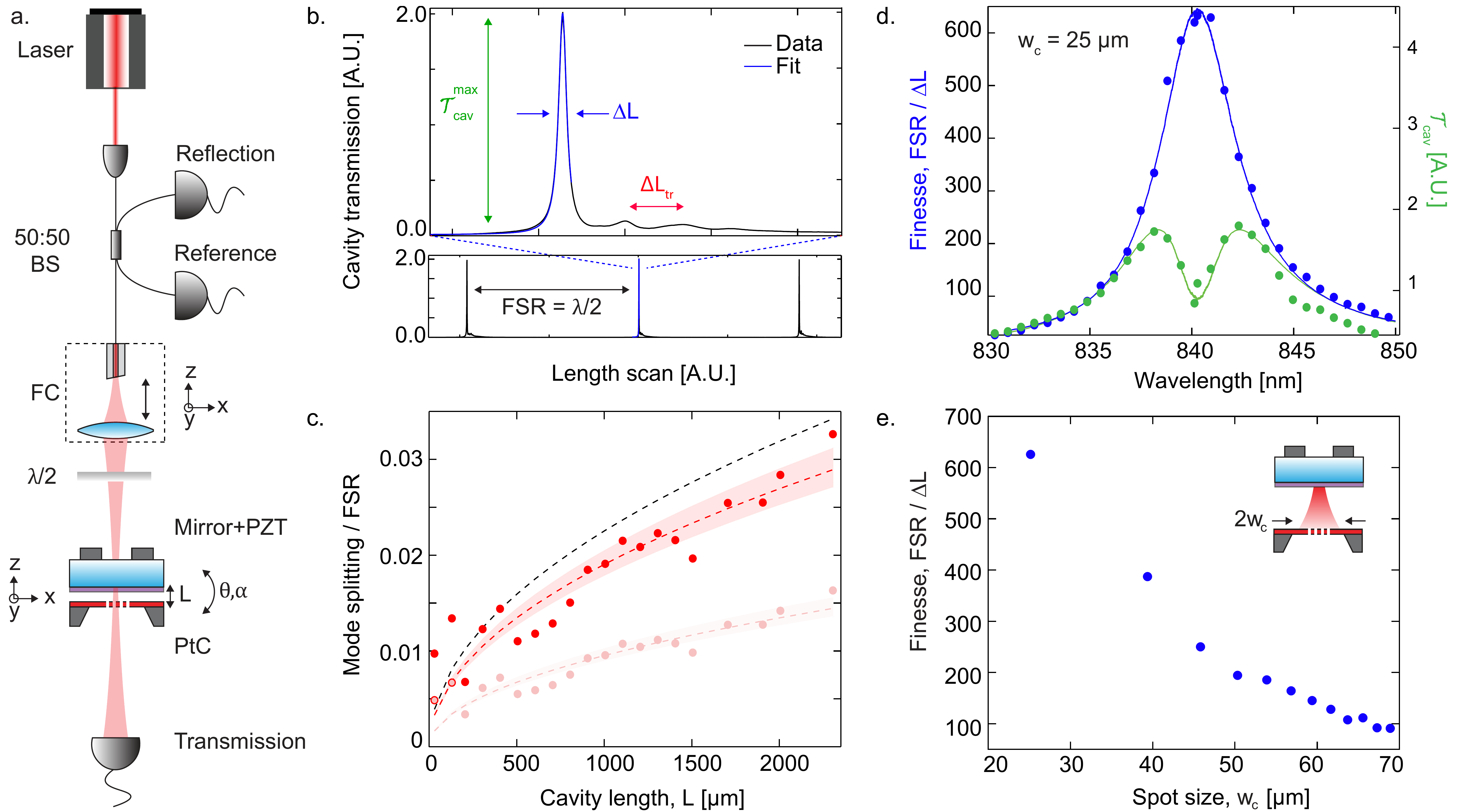}
\caption{Cavity-based transmission measurements. (a) Setup for probing a hemispherical Fabry-Perot cavity based on focusing metamirror. (b) Bottom panel: Measured cavity transmission  $\mathcal{T}_\t{cav}$ (black) versus cavity length ($L$), scanned over multiple free spectral ranges ($\t{FSR}=\lambda/2$). Top panel: $\mathcal{T}_\t{cav}$ near resonance, highlighting the linewidth $\Delta L$ and peak transmission $\mathcal{T}_\t{cav}^\t{max}$ of the fundamental mode, and transverse mode splitting $\Delta L_\t{tr}$. A Lorentzian fit (blue) yields a finesse of $\mathcal{F}=0.6\times 10^3$. (c) Measured $\Delta L_\t{tr}$ versus cavity length.  Red (pink) points assume no (one) missing peak between adjacent pairs of peaks as in panel b.  Dashed lines are fits to Eq. \ref{eq:modesplitting} for $2f=$ 20 cm (black), 28 cm (red), and 112 cm (pink). Shaded regions are 95\% confidence bounds. (d) Measured $\mathcal{F}$ (blue dots) and $\mathcal{T}_\t{cav}^\t{max}$ (green dots) versus wavelength for a $L=30 \upmu$m cavity. The blue curve is inferred from the free-space transmission measurement in Fig. \ref{fig:freespaceMeasurements}b.  The green curve is the corresponding transmission model (Eq. \ref{eq:cavityTransmission}), assuming $\mathcal{L} = 0.85\%$. (e) Measured cavity finesse versus inferred mode radius ($w_\t{c}$), corresponding to $L = [0.03,1.5]\,\t{mm}$.}
\label{fig:cavityMeasurements}
\vspace{-2mm}
\end{figure*} 

As a base case, we first carried out free space measurements on a planar hexagonal Si$_3$N$_4$ membrane mirror.  Our experimental setup is shown in Fig. \ref{fig:freespaceMeasurements}a and features a tunable diode laser (Newport Velocity) capable of mode-hop-free wavelength scanning from $830-852$ nm, calibrated using an independent wavemeter.  For free-space measurements, the laser is passed through an optical fiber and then focused onto the mirror, using a gimbal mount and translation stages to tune alignment via retroreflection into the fiber.  By changing the fiber-coupler collimation (and translating the sample to the new beam focus), we vary the spot size on the mirror from $25 - 100\,\upmu\t{m}$.  The quantity we measure is the ratio of transmitted to incident power---the free-space transmissivity $\mathcal{T}_\t{fs}$---whose relation to the mirror reflectivity can be modeled~as
\begin{equation} \label{eq:Tfs}
   \mathcal{T}_\t{fs} = 1 - \mathcal{R} - \mathcal{L}(1-\eta) = \mathcal{T}+\mathcal{L}\eta
\end{equation}
where $\mathcal{L}$ is the mirror loss (e.g., due to scattering and absorption), $\eta$ is the fraction of loss collected by the transmission detector, and $\mathcal{T} = 1-\mathcal{R}-\mathcal{L}$ is the mirror transmissivity.

Transmission versus wavelength measurements for the plano mirror, shown in Fig. \ref{fig:freespaceMeasurements}b, corroborate the unit cell simulation with a slightly shifted center wavelength. Specifically, for a narrow range of wavelengths centered on 844 nm, we observe transmission as low as $\mathcal{T}_\t{fs} = 0.1\%$, with off-resonant behavior in good agreement with simulated $\mathcal{T}$ assuming a membrane thickness of 185 nm.  (We note that the apparent reflectivity $\mathcal{R}\approx 99.9\%$ is on par with 1550 nm hexagonal Si$_3$N$_4$ PtC metamirrors on which our unit cell geometry is based \cite{zhou2022cavity}.) 
As highlighted in Fig. \ref{fig:freespaceMeasurements}c, we observe that the $\mathcal{T}_\t{fs}$ depends strongly on spot size, decreasing with smaller spot sizes.  This behavior is partially consistent with clipping loss due to the finite mirror aperture, as discussed below. 

Free-space transmission measurements on our focusing metamirror are shown in Fig. \ref{fig:freespaceMeasurements}d-e, revealing behavior similar to the planar mirror but with a $5$ nm lower center wavelength and $10$-fold larger transmission.  For these measurements, special care was taken to laterally and longitudinally center the beam by using the mirror perimeter as a knife edge and by retroreflecting off the membrane outside the focusing mirror, respectively.  The lowest measured transmissivity ($\mathcal{T}_\t{fs}\approx 1\%$ at a spot size of $25\,\upmu\t{m}$) is 5 times higher than the intensity-averaged mirror transmissivity inferred from the blue model curve in Fig. \ref{fig:mirrorDesign}c, suggesting that $\mathcal{T}_\t{fs}$ may include a significant loss component in addition to clipping.

We next turned our attention to the cavity-based characterization of the focusing metamirror.  As shown in Fig. \ref{fig:cavityMeasurements}a, the experimental setup is identical, except the metamirror is preceded by a plano dielectric mirror mounted on a 5-axis alignment stage, forming a FP.  We use a custom dielectric coating (ATFilms) with $\mathcal{T}_0<0.01\%$ at 850 nm \cite{wilson2012cavity} to ensure that the cavity is heavily overcoupled in transmission.  A piezo is used to scan the cavity length $L$ over several free spectral ranges.  Focusing first on the fundamental transverse cavity mode, the quantities we measure are the linewidth $\Delta L$ and peak value $\mathcal{T}_\t{cav}^\t{max}$ of the cavity transmission $\mathcal{T}_\t{cav}(L)$, from which metamirror transmissive properties can be inferred according to the model for a low-loss $(1-\mathcal{R}\ll 1$), overcoupled ($\mathcal{T}_0\ll \mathcal{T}$) FP

\begin{subequations}\label{eqs:cavityFinesseTransmission}
\begin{equation}\label{eq:cavityFinesse}
    \frac{\lambda}{2\Delta L} \equiv \mathcal{F} \approx  \frac{2\pi}{1-\mathcal{R}} = \frac{2\pi}{\mathcal{T}+\mathcal{L}}
\end{equation}
\begin{equation}\label{eq:cavityTransmission}
    \mathcal{T}_\t{cav}^\t{max}(\mathcal{L}) = \frac{\mathcal{T}}{(\mathcal{T}+\mathcal{L})^2} 4\mathcal{T}_\t{0} =  \frac{2\pi/\mathcal{F}-\mathcal{L}}{(2\pi/\mathcal{F})^2} 4 \mathcal{T}_0
\end{equation}
\end{subequations}

where $\mathcal{F}$ is the cavity finesse.  In addition, we measure the splitting $\Delta L_\t{tr}$ between adjacent higher-order cavity modes.  Assuming the metamirror behaves the same as a concave dielectric mirror, $\Delta L_\t{tr}$ yields its focal length, according to \cite{kogelnik1966laser}
\begin{equation} \label{eq:modesplitting}
\Delta L_\t{tr} = \frac{\lambda}{2\pi}\cos^{-1}\sqrt{1-\frac{L}{2f}}
\end{equation}

A representative cavity length sweep is shown in Fig. \ref{fig:cavityMeasurements}b, revealing both the reflectivity and focal length of the metamirror via the cavity linewidth and transverse mode-splitting, respectively.  For finesse measurements, we fit the fundamental peak to a Lorentzian and normalize to the free spectral range between mode families as shown in the bottom panel.  For transverse mode-splitting measurements, we average the distance between adjacent peaks within mode families.

A key result of our study is presented in Fig. \ref{fig:cavityMeasurements}c, in which the cavity mode-splitting $\Delta L_\t{tr}$ is plotted as a function of cavity length in the range $L = [0.03,2.3]\,\t{mm}$. Compared to the extended mirror model in Eq. \ref{eq:modesplitting}, we find that the observed splitting is consistent with a radius of curvature of $2f = 28\,\t{cm}$.  We emphasize that this inference assumes the absence of hidden peaks in cavity spectra as in Fig. \ref{fig:cavityMeasurements}a---for example, if a single hidden peak resides between each pair, then the actual mode splitting would be half as large. This possibility is depicted by the pink points in Fig. \ref{fig:cavityMeasurements}c, and would suggest $2f= 112\,\t{cm}$. (It could likewise account for the discrepancy between the models at the two shortest cavity lengths.)  Nevertheless, to our knowledge, the observed scaling of $\Delta L_\t{tr}$ with $L$ constitutes the first demonstration of a suspended, concave metamirror.

Measurements of cavity finesse and peak transmission are shown in Fig. \ref{fig:cavityMeasurements}d-e, and corroborate the free space measurements augmented by a parasitic loss at the level of $\mathcal{L}\approx1\%$.  Specifically, in Fig. \ref{fig:cavityMeasurements}e, we plot finesse versus spot size by varying the cavity length and assuming the cavity mode waist is that of a hemispherical cavity $w_c \approx (2 L f \lambda^2/\pi^2)^{0.25}$ with $f = 14\;\t{cm}$ (Fig. \ref{fig:cavityMeasurements}c). The higher finesse at small inferred $w_\t{c}$ (small $L$) agrees quantitatively with the free-space measurements in Fig. \ref{fig:mirrorDesign}a, and reaches a maximum of $\mathcal{F} = 0.63\times 10^3$ at $w_\t{c} = 25\,\upmu\t{m}$ $(L = 30\,\upmu\t{m})$, corresponding to $\mathcal{T}+\mathcal{L} \approx 1\%$. To estimate the loss contribution at this spot size, as shown in Fig. \ref{fig:cavityMeasurements}d, we compared measurements of  $\mathcal{T}_\t{cav}^\t{max}$ and $\mathcal{F}$ versus laser wavelength.  At large detuning from the center wavelength $\lambda_0$, we found that $\mathcal{T}_\t{cav}^\t{max}/\mathcal{F}$ was roughly constant, consistent with $\mathcal{T}\gg \mathcal{L}$ in Eq. \ref{eqs:cavityFinesseTransmission}.  Near $\lambda_0$, the drop in normalized transmission to $\mathcal{T}_\t{max}(\mathcal{L})/\mathcal{T}_\t{max}(0) \approx 0.15$ is consistent with $\mathcal{L}\approx 0.85\%$, suggesting the cavity is near critically coupled.

\vspace{-2mm}
\section{Discussion - Loss and Radius}
\vspace{-2mm}

\begin{figure}[t!]
	\vspace{-2mm}
	\centering
	\includegraphics[width=0.9\columnwidth]{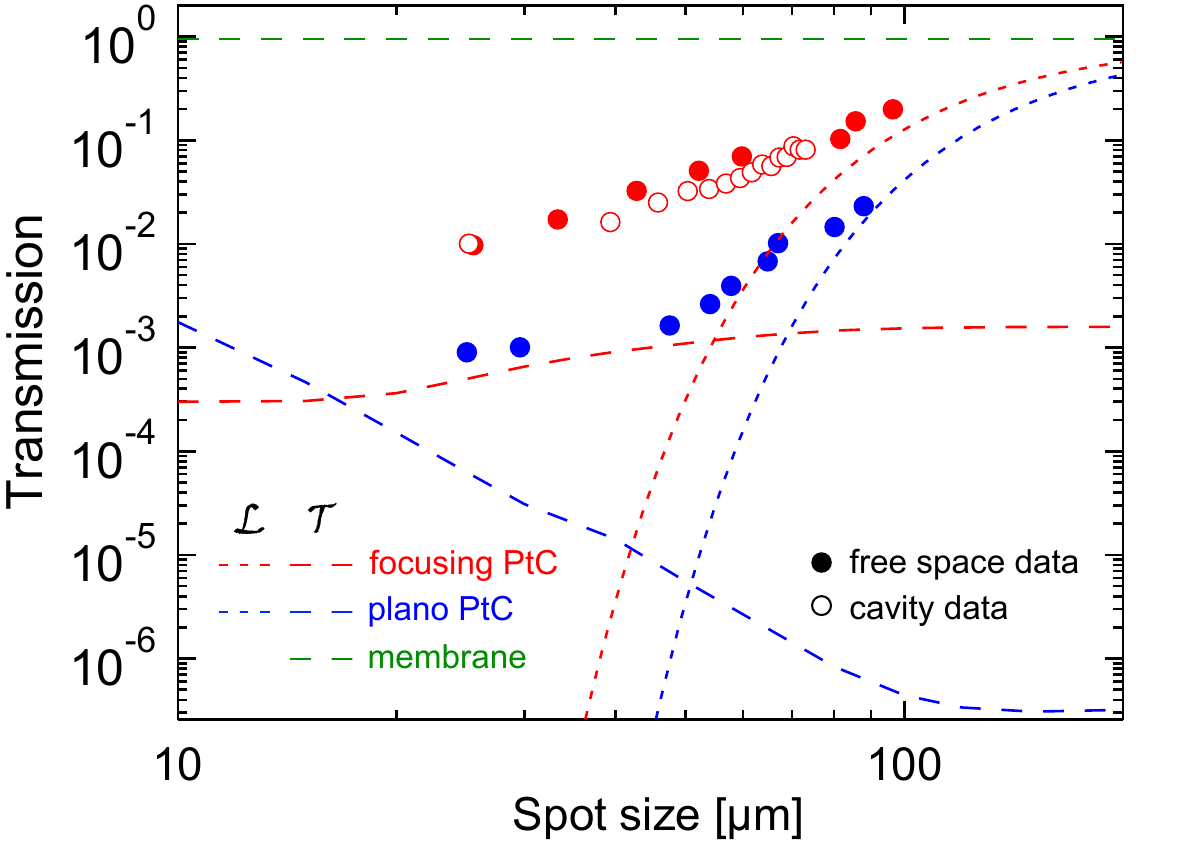}
	\caption{Measured transmission of plano (blue) and focusing (red) metamirrors compared to models clipping loss (short dash), photonic crystal transmission (long dash), and membrane transmission (green dashed). Solid and open circles correspond to free space and cavity-based transmission measurements, respectively.}
    \label{fig:losses}
	\vspace{-2mm}
\end{figure} 

We conclude by speculating on the source of two discrepancies between our design and characterized focusing metamirrors: parasitic loss $\mathcal{L}$ and the larger measured focal length.

With regards to the source of parasitic loss $\mathcal{L}$, two observations offer clues:  First, in both our free space and cavity-based measurements, the apparent transmission---$\mathcal{T}_\t{fs}$ and $2\pi/\mathcal{F}$, respectively---increases with spot size $w$; moreover, their magnitudes agree, consistent with forward-scattered loss ($\eta = 1$ in Eq. \ref{eq:Tfs}). Second, in our free space measurements, both planar and focusing metamirrors exhibit similar $\mathcal{T}_\t{fs}$ versus $w$ scaling, but with a 10-fold lower $\mathcal{T}_\t{fs}$ for the planar mirror.  These observations appear to rule out absorption (inconsistent with $\eta = 1$) but leave open the possibility of scattering, diffraction, or clipping loss.  They could also be explained by a spot-size-dependent mirror transmission $\mathcal{T}(w)$ due to the angle-dependence of the unit cell reflectivity, the inhomogeneous reflectivity of the aperiodic PtC, or a combination thereof.

To investigate the loss mechanism further, in Fig. \ref{fig:losses}, we compare models of clipping and transmission loss to free-space and cavity-based transmission measurements of plano and focusing PtC mirrors.  For clipping loss, we use a circular aperture model $\mathcal{L}\approx T_\t{mem} e^{-2 w_m^2/w^2}$, where $T_\t{mem}$ is the bare membrane transmissivity.  For transmission through the planar metamirror, we average the simulated unit cell plane wave transmissivity $\mathcal{T}(\vec{k})$ over the plane wave distribution $P(\vec{k},w)$ of a Gaussian beam, where $\vec{k}$ is the incident propagation vector. 
 For transmission through the focusing mirror, we average the design unit cell transmissivity $\mathcal{T}(x,y)$ (Fig. \ref{fig:mirrorDesign}c) over the Gaussian beam intensity profile $I_0 e^{-2(x^2+y^2)/w^2}$.  Evidently, the clipping loss is consistent with both free space measurements at large spot size, assuming a (measured) membrane transmissivity of $T_0\approx 94\%$ and mirror radii of $w_m = 125(100)\,\upmu\t{m}$ for the plano (focusing) mirror. At smaller spot sizes, plano and focusing mirror transmissivities appear to asymptote to $0.1\%$ and $1\%$, respectively. Both are roughly an order of magnitude above modeled transmissivities, suggesting additional sources of loss and/or transmission non-ideality.  
 The significantly lower transmission of focusing mirrors at small spot sizes, inferred to be parasitic loss by the cavity finesse measurement in Fig. \ref{fig:cavityMeasurements}d, seems plausibly related to the coarse (30-step) discretization of the non-periodic unit cell profile, evident in Fig. \ref{fig:mirrorDesign}c.  It may also be due to the deformation of the PtC upon release of the membrane, as discussed below.
 
With regards to the discrepancy between the design ($f=10\,\t{cm}$) and measured ($f\approx 14\,\t{cm}$) focal length, we note that applying in-plane stress to a focusing PtC metamirror is an established method to modify its focal length \cite{cheng2019mechanically} via the associated deformation of the PtC geometry. Since our Si$_3$N$_4$ membrane is under ${\sim} 1\t{GPa}$ tensile prestress due to thermal mismatch with the Si substrate, it will relax upon release, resulting in a non-uniform change of the unit cell size depending on its fill factor and radial position.  We speculate that this may account for the discrepancy between the design and measured focal length and that accounting for stress relaxation will be important for future refinement of the PtC design.

\vspace{-3mm}
\section{Summary and Outlook}
\vspace{-3mm}

In summary, we have used non-periodic PtC patterning to realize a high-reflectivity, focusing metamirror in a 200-nm-thick Si$_3$N$_4$ membrane. The metamirror reflectivity $\mathcal{R}\approx 99\%$ and focal length $f \approx 14\;\t{cm}$ were confirmed by constructing $30-2300\;\upmu\t{m}$ long Fabry-P\'{e}rot cavities in tandem with a plano dielectric mirror, observing a critically coupled finesse as high as $0.63\times 10^3$, and observing transverse mode splitting consistent with that of a plano-concave cavity. A peculiar feature observed in both free space and cavity-based measurements is an apparent decrease of total mirror loss $\mathcal{T}+\mathcal{L}$ at smaller spot sizes.  The magnitude at small spot sizes is inconsistent with clipping loss and unit cell transmission models, and may be inherent to the PtC nonperiodicity and, or, stress relaxation of the membrane upon release \cite{cheng2019mechanically}. 

In future work, we envision refining the PtC design to reduce discretization and compensate for stress relaxation.  Another goal is to integrate focusing metamirrors into a phononically engineered (e.g., trampoline \cite{reinhardt2016ultralow}  or phononic crystal \cite{tsaturyan2017ultracoherent}) membrane to realize a high-cooperativity, vertically-integrated cavity optomechanical system \cite{guo2017integrated}.  We emphasize that a unique feature of vertical integration is its compatibility with high-aspect-ratio membranes with optionally large $Q/\t{mass}$ or $Q\times\t{mass}$ factors.  It is also compatible with wafer-scale cavity arrays, free space readout, and or integrated electrodes. These options give access to a diversity of applications ranging from optomechanical inertial sensing \cite{chowdhury2023membrane} and force microscopy \cite{halg2021membrane}, to entanglement-enhanced distributed sensing \cite{xia2023entanglement}, to hybrid quantum electromechanical systems \cite{andrews2014bidirectional}.

\vspace{-3mm}
\section*{Acknowledgments}
\vspace{-3mm}

The authors thank Samuel Deléglise, Thibaut Jacqmin, Himanshu Patange, and John Lawall for invaluable discussions about photonic crystal design and fabrication.  We also thank Christian Pluchar and Mitul Dey Chowdhury for helping design the optical characterization setups and Ewan Wright for helping with numerical modeling. This work is supported by NSF Grants ECCS-1945832 and PHY-2209473. A. R. A. acknowledges support from a CNRS-UArizona iGlobes fellowship. Finally, the reactive ion etcher used for this study was funded by an NSF MRI grant, ECCS-1725571.

\bibliography{ref}
\end{document}